# ACOUSTIC FEATURES AND PERCEPTIVE CUES OF SONGS AND DIALOGUES IN WHISTLED SPEECH: CONVERGENCES WITH SUNG SPEECH

**Julien Meyer**

*Laboratory of Applied Bioacoustics, Universitat Politechnica de Catalunya, Avda Rambla Exposició s/n 088000 Vilanova i la Geltrú, Barcelona, Spain*
*Julien.Meyer@univ-lyon2.fr*

**Abstract**

Whistled speech is a little studied local use of language shaped by several cultures of the world either for distant dialogues or for rendering traditional songs. This practice consists of an emulation of the voice thanks to a simple modulated pitch. It is therefore the result of a transformation of the vocal signal that implies simplifications in the frequency domain. The whistlers adapt their productions to the way each language combines the qualities of height perceived simultaneously by the human ear in the complex frequency spectrum of the spoken or sung voice (pitch, timbre). As a consequence, this practice underlines key acoustic cues for the intelligibility of the concerned languages. The present study provides an analysis of the acoustic and phonetic features selected by whistled speech in several traditions either in purely oral whistles (Spanish, Turkish, Mazatec) or in whistles produced with an instrument like a leaf (Akha, Hmong). It underlines the convergences with the strategies of the singing voice to reach the audience or to render the phonetic information carried by the vowel (tone, identity) and some aesthetic effects like ornamentation.

## INTRODUCTION

The pronunciation of spoken or sung words has been naturally adapted into whistles in several cultures. This complementary form of speech is used to overcome ambient noise and fight reverberation when the speaker is far from his interlocutor or from his audience. Its first functional use is to guarantee a good intelligibility of sentences in conditions for which spoken dialogues would be inefficient. Moreover, it is popular in some cultures for its aesthetical qualities and is therefore used to emulate songs. Such a traditional oral behaviour has been recently documented as still surviving in at least 14 languages all over the world: in Africa (Ewe, Ari), in Asia (Akha, Hmong), in America (Chinantec, Gaviao, Mazatec, Mixtec, Siberian Yupik, Surui), in Europe (Greek, Spanish, Turkish), in Oceania (Abu-Wam) [Meyer, 2005].

To emulate in whistles a sentence of his language, the whistler encodes the words in a modulated narrow band of frequency [Busnel and Classe, 1976]. The main acoustic transformation is therefore in the frequency domain: from the multidimensional frequency spectrum of the voice to the mono-dimensional one of whistles. At the perceptive level, the linguistic information is synthesized in a single whistled pitch. On the contrary, the signal of the voice carries two perceptive qualities of height that are combined differently in each language to encode the words. These two qualities are the ones that have been observed in several perceptive tests in musical acoustics: (i) the perceptive sensation resulting from the F0, called pitch; and (ii) the perceptive sensation resulting from the complex aspect of the frequency spectrum, called timbre in music [Risset, 1968]. For example, pitch plays a primordial role in the intelligibility of several tone languages and the timbre space strongly characterizes the quality and the identity of a vowel through the formants. In response to these aspects, the whistled practices have adapted to the phonological rules of organisation of sounds of the transposed languages, selecting salient cues to optimise the intelligibility of the receiver.

The present paper underlines the convergence of some aspects of whistled speech with the singing voice while providing an overview of this phenomenon. A particular attention is drawn to the practice of popular singing and to the diversity of the concerned languages. Most of the whistled data analyzed here was documented since 2003 during fieldwork projects associating local researchers.

## ACOUSTIC ADAPTATATION TO CONSTRAINTS

**Production**

In whistling and singing, mastering the technique of production requires a phase of learning. An efficient emission relies on a homogenous, powerful, relaxed and precise control of both the air flow and the physiological constraints imposed by such types of articulations of words. In comparison to speaking, both increase the tension of the muscles of the vocal tract. These tensions reinforce the concentrations of energy in the signal because, apart from aesthetical and intelligibility exigencies that whistled and sung speech share in some cases, both have the common aim to carry the oral sound in the distance. As a sound source, the singing voice uses the vocal chords and a vocal tract often modified by a low larynx and large pharynx. The resulting signal bears a complex frequency spectrum characteristic of the human voice and largely described in the literature. For whistling, the lips are tight and the sound is produced by the turbulence of the air forced either into the smallest hole of the vocal tract or against an edge (depending of the technique). The reduced mouth cavity acts as a resonator to tune the sounds (figure 1). In some cultures a leaf can be used to whistle a message with a linguistic attitude, in these cases, the properties of vibration this instrument also play an important role (figure2).

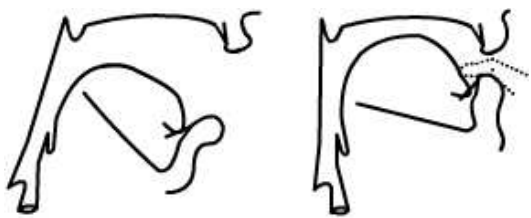

*Figure 1: Open vocal tract. While singing (left) and while whistling (right)*

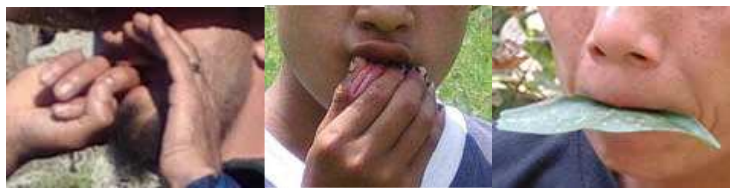

*Figure 2: A selection of techniques of whistling used to emulate speech*

**Strategies to reach the audience**

The acoustic strategies of whistling and singing share a common behavior in a significant band of frequency. The singing voice of a trained singer shows a concentration of energy between 2 and 4 kHz, especially for classical operatic singers of western music. It is formed by the proximity of two or three of the formants 3, 4, 5 or by the amplification of one of the formants already situated in this frequency domain. It has been called the *Singing Formant* and is sufficiently compact to favour a perceptive integration of the concerned frequencies [Bartholomew 34, Sundberg 1972, Sundberg 2001]. In parallel, all whistled forms of languages use functional frequencies of whistling between 1 kHz and 4 kHz. In both cases, two reasons justify the choice of these bands of frequencies: firstly they place the signal above most of the concurrent frequencies of the performers´ environment; secondly they cover a domain that corresponds to the best performances of hearing in terms of selectivity and audibility. As a consequence the signal to noise ratio (SNR) at the reception is sufficiently high for a good perception. For example, in the case of singing, Pierce [1983; cited by Woisard-Bassols, 2001] showed that the perception of the singer's voice above a chorus or an orchestra relies essentially on the singing formant. He measured that the SNR at its peak (2500 Hz) for a tenor voice was around 30 dB. Similarly, in the case of whistled speech, the signal remains largely above the natural background noise at relatively long distance. For example: an equivalent SNR of 30 dB -SNR measured in the frequency band of whistles- is obtained at around 200 m of the whistler for a sentence produced at 90-100 dB in a quiet environment (around 40 dB of noise). Apart from this common attitude, singing and whistling also show proper characteristics due to their sound source: sung speech shows amplifications around formant 1, whereas the whistled signal is shaped as a natural telecommunication system. Indeed, the bandwidth of its fundamental frequency emerging from the noise is less than 500 Hz, and its dynamic in amplitude is reduced compared to spoken speech [Meyer and Gautheron 2005]. Contrary to what occurs in the singing voice, the harmonics do not group in formants and do not play a fundamental role in speech recognition.

## PHONETIC STRATEGIES: CONVERGENCES

**Lengthening of vocalic duration**

When compared to the spoken form of speech, the vocalic duration in the whistled form of speech is increased. Differences occur in function of the distance of the interlocutor. I measured that, for a dialog made at 150 m of distance, the vowels lasted meanly 26 % more in whistled Turkish than in spoken Turkish and 28 % more in whistled Akha than in spoken Akha of North Thailand. Some differences may appear in function of the individuals, as shown by Moles [1970] in whistled Turkish as the speakers keep their characteristics of elocution during the whistled articulation. At very long distances of practice (several km) or for the sung mode of whistled speech, the lengthening of the vowels can reach meanly 50 %; with some vowels much more emphasized than others. Contrary to what occurs in singing, such exaggerated durations do not reduce the intelligibility but still ease it. Indeed, the vowels with a very long duration are most of the time situated at the end of a speech group; this way, they help to sequence rhythmically a sentence in coherent units of meaning.

The sung form of speech is also characterized by an increase in vowel duration in most of the singing styles. It is the result of the combination of many parameters influencing the production in singing voice. First, artistic choices justify this strategy, for example to control the homogeneity of the timbre quality, especially in the *bel canto* technique [Woisard-Bassols, 2001]. Moreover, a slower tempo is often inherent to music and results in a slower rate of speech. Then, the articulatory movements last longer because of their greater range of variation [Scotto Di Carlo, 2005b]. Finally, this phonetic attitude partly relies on the fact that it is efficient to ease the perception of the voice quality and therefore favour a comfortable listening. In a similar way to what occurs in shouted and whistled speech, vowel lengthening is also an adaptation to the difficult conditions of communication. Within certain limits this phenomenon increases intelligibility; yet, when the duration of the vowels is beyond a tolerance threshold, the performances of recognition are affected and decrease [Scotto di carlo, 2005a].

**Influence of the language phonology**

The case of whistled speech

As explained in introduction, among the several languages of the world that have been compared according to their whistled behaviour, different types have been highlighted [Meyer 2005]. In most of the tone languages (like Mazatec, Chinantec, Akha, Hmong) whistling selects primarily pitch cues carried by tone registers and tone contours. In most of the non tonal languages (like Greek, Turkish, Spanish) it selects primarily timbre elements carried by segmental cues of the voice; and in an intermediary category of languages it selects cues from both pitch and timbre by jumping in real time for one to the other (tonal Surui, non tonal Siberian Yupik and Chepang). In fact, for each language the whistlers give priority in frequency to a dominant trait that is carried in the spoken voice either by the *formant distribution* (most of the non tonal languages; figure 3, left side) or by the fundamental frequency (most of the tonal languages; figure 3 right side). But in the case of an incipient tone language like Chepang or a language using a lot stress like Siberian Yupik, the contribution of both is balanced, which explains their intermediate strategy in whistles. The reduction of the frequency space in whistles classifies therefore the languages in typological categories. These categories depend on the respective roles played by the Fo and the formant distribution in the intelligibility of the spoken form. For example, in a tone language, variation in pitch or its physical correlate, fundamental frequency (F0), is used to contrast word meaning. Sometimes its contribution to intelligibility is greater than vowel identity, like in Chinese (17% for tones vs. 12 % for vowels [Fu and al 1997]) whereas in other languages like the two-tone Surui of Amazon, it is not enough to encode a large vocabulary of words. On the contrary, in several non tonal languages, the information encoded in the Fo -like stress- plays a secondary role for intelligibility whereas the aspects encoded in the formants are primordial (vowel identity and consonant transitions). The typology of whistled form of speech available in Meyer [2005] underlines that the specificities of each language are mirrored in whistled speech.

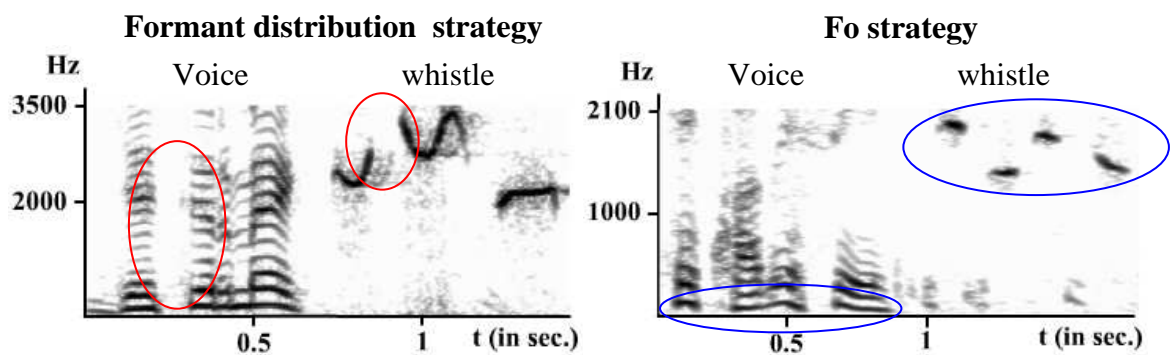

*Figure 3: Left, a Turkish word spoken and whistled (/getirmek/). The /t/ in indicated. Right, a word spoken and whistled - pitch transposition- in Mazatec (4 tone language).*

Comparison with sung speech

In a similar way to what occurs in whistled speech, the contributions for encoding the word meaning of both the Fo and the *formant distribution* may influence the composition and the interpretation of songs. Yet, due to the complex frequency spectrum of the singing voice, testing this aspect is less simple. Furthermore, other parameters enter in consideration: sung speech is not only focused on intelligibility as it takes into account artistic priorities that might interfere with the phonetics and the identification of words. For example, in operatic singing a soprano voice is much less intelligible than a bass [Scotto di Carlo 2005b]. Moreover, within a language there may be a wide range of variation between different styles of songs, some of them been borrowed from other cultures.

For the singing voice in non tonal languages, the research has been logically focused on understanding the aspects of the formant distribution influencing intelligibility [Gottfried and Chew, 1986; Scotto di Carlo 2005b]. One of the main results is that the higher is the pitch, the more difficult is the identification of the phonemes. Moreover, the difficulty to identify the vowels in sung speech has been also attributed to the fact that the formants 2 and 3 are less phoneme-dependent than in spoken speech. These results underline that the intelligibility is advantaged when the frequency spectrum shows compact formants with a phoneme-dependent distribution. Such a conclusion is consistent with the fact that the human perception is sensible to formant proximity, especially for identifying the vowels [Chistovitch and Lublinskaja, 1979; Schwartz and Escudier, 1989]. For whistled speech it is an important aspect as such proximities would explain the frequency distribution of whistled vowels in non tonal languages [Meyer, 2005], with /i/ among the highest frequencies, /o/ among the lowest and /e/ and /a/ in between (figure 4). It would explain also the good performances of non whistlers of such languages to identify whistled vowels, /i/ being perceived as an acute timbre in spoken speech, and /o/ as a low one [Meyer and al, 2007].

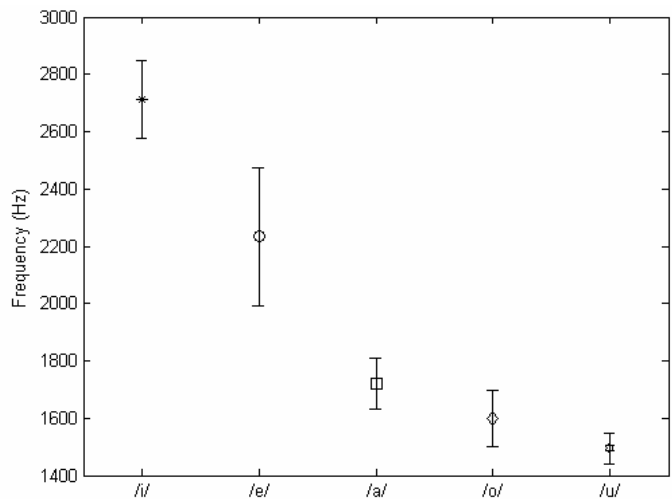

*Figure 4: Frequency distribution of whistled Spanish vowels*

In the case of tone languages, the Fo parameter of singing voice is easy to test. For example, Wong and Diehl [2002] investigated songs of Cantonese to further understand how songwriters reconcile the possible linguistic role of $F_0$ variation in expressing a lyric with the musical role of $F_0$ variation in specifying melodic intervals. According to their preliminary enquiry three options appear: *"The first is to ignore lexical tones and word meaning and to use pitch exclusively to mark the melody. […] The second option is just the reverse: to preserve the normal pitch variations of lexical tones while ignoring the melody, sacrificing musicality for intelligibility.[…] The third option is intermediate between the first two. […] This preserves musicality at the cost of reduced lyric intelligibility."* [Wong and Diehl 2002, p. 203]. These three attitudes have been observed in several studies: (i) the first option preserves musicality at the cost of reduced lyric intelligibility. It concerns mostly fixed melodies imposed to lyrics. Chao [1956] reported having observed it in Chinese contemporary songs. Similarly, Saurman [1996] measured the direction of pitch change over consecutive syllables in tones and song pitches: they matched in only 32 % of the cases for the Thai national anthem and 42 % for a foreign hymn translated into Thai. In Mexico, Casimiro [2007, pers. com.] reported the same phenomenon for the Mexican national anthem sung in Mazatec language at school. (ii) In the case of the second option, songs are very speech-like. It is the case of singsong themes in Chinese described by Chao [1956]. Moreover, according to the description of Van-Khe [1997] this option generally concerns a large body of traditional popular songs in Vietnam and of several cultures of South East Asia: *"In the popular traditions of South-East Asia the improvisation is mostly poetic. The melody of the songs must follow the linguistic intonation of their text otherwise it would change the poetic signification"* [Van-Khe, 1997]. Saurman [1996] confirmed this assertion for traditional Thai songs. I measured that it is also the case for popular songs in Akha language of North Thailand. Classical styles of singing are also concerned by this option. For example, Yung [1983] found a melody-tone relationship in Cantonese opera, that was somewhat similar to what Chao [1956] reported in singsongs. (iii) In the case of the third option: *"songwriters may attempt to preserve at least partially the pitch contrasts of lexical tones while not unduly restricting the melodic role of F0 changes."* [Wong and Diehl, 2002, p. 203]. Wong and Diehl proved that this strategy was at play in several contemporary songs in Cantonese. They not only examined the strategies used by composers and singers to reconcile linguistic and musical roles of F0, they also proved that corresponding strategies are used by listeners to extract the lexical meaning of the lyric: listeners apply an ordinal mapping rule in assigning tone categories.

## SUNG MODE OF WHISTLED SPEECH: A FIRST APPROACH

The sung mode of whistled speech is the practice that consists of emulating the sung voice while producing a whistled sentence. It is not an aspect extensively developed in all the cultures concerned by whistled speech. It seems much more common in tone languages. It must be due to the fact that such languages often reconcile melody and meaning in singing them through a same physical parameter (Fo, see preceding paragraph). Usually, repertoires of old popular songs are concerned by this practice. Yet it is sometimes adapted to contemporary songs.

As an example, I present the Akha sung speech whistled with a leaf. Akha is a three tone language of south-east Asia for which whistled speech transposes primarily the Fo. It has a very rich oral repertoire of songs. Some of them, such as love songs, are more often used with whistling because the sung leaf is popular for courtship. As the aim is to transmit a poetic message, the melody of the songs follows the pitch of their lyrics. The acoustic propagation of this type of signal enables the message to reach easily the lover inside his/her wooden house (physical justification). The whistlers also say that whistling is more melodic (aesthetical justification). Indeed, several acoustical features are consistent with this second justification. First, whistling with an instrument like a leaf gives more sound energy to the harmonics (figure 5). Then the singers use classical techniques of ornamentation: they combine vowel lengthened durations with a vibrato. Finally, the singers impose at their sung sentence a modified rhythm: for example the rhythm is often iambic (stress is marked by a lengthened duration). The tone perception is not perturbed by the oscillations produced by the vibrato as these do not transcend the median frequency of the interval separating two tones (the tones of a same sentence do not overlap even in vibrato, figure 5).

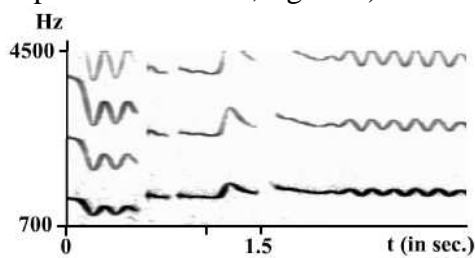

*Figure 5: Extract of an Akha song whistled with the leaf, with ornamentations.*

## CONCLUSIONS

Whistled and sung speech share several acoustic, phonetic and articulatory properties underlined in this paper. Such a comparative approach has also detailed how the phonology of a language influences both sung and whistled adaptations of speech. As the research on the singing voice has a long history, several phonetic results in this domain contribute to better understand the strategies of transformation of the spoken voice into whistles. For example, they provide an alternative point of view on the perception of the formant distribution, which is primordial to whistle a non tonal language. Finally, in some cultures sung and whistled speech merger in a one phenomenon that was called the *sung mode of whistled speech*. All these aspects have been illustrated thanks to unprecedented data collected in the field during the last four years.